\documentclass[prd,twocolumn,tightenlines,showpacs,nofootinbib]{revtex4-1} 

\usepackage{amsfonts}
\usepackage{dcolumn}
\usepackage{bm}
\usepackage{amsmath}
\usepackage{nicefrac}
\usepackage{amsthm}
\usepackage{amssymb}
\usepackage{graphicx}
\usepackage{color}
\usepackage{url}

\newcommand{\appropto}{\mathrel{\vcenter{
  \offinterlineskip\halign{\hfil$##$\cr
    \propto\cr\noalign{\kern2pt}\sim\cr\noalign{\kern-2pt}}}}}
\renewcommand{\v}[1]{\boldsymbol{#1}}		

\begin{document}

\title{Enhanced effects of variation of the fundamental constants in laser interferometers and application to dark matter detection}

\date{\today}
\author{Y.~V.~Stadnik$^1$} 
\author{V.~V.~Flambaum$^{1,2}$} 
\affiliation{$^1$ School of Physics, University of New South Wales, Sydney 2052, Australia}
\affiliation{$^2$ Mainz Institute for Theoretical Physics, Johannes Gutenberg University Mainz, D 55122 Mainz, Germany}

\begin{abstract}
We outline new laser interferometer measurements to search for variation of the electromagnetic fine-structure constant $\alpha$ and particle masses (including a non-zero photon mass). 
We propose a strontium optical lattice clock -- silicon single-crystal cavity interferometer as a novel small-scale platform for these new measurements. 
Our proposed laser interferometer measurements, which may also be performed with large-scale gravitational-wave detectors, such as LIGO, Virgo, GEO600 or TAMA300, may be implemented as an extremely precise tool in the direct detection of scalar dark matter that forms an oscillating classical field or topological defects. 
\end{abstract}



\maketitle 


\section{Introduction}
Dark matter remains one of the most important unsolved problems in contemporary physics. Astronomical observations indicate that the energy density of dark matter exceeds that of ordinary matter by a factor of five \cite{Bertone2010Book}. Extensive laboratory searches for weakly interacting massive particle (WIMP) dark matter through scattering-off-nuclei experiments have failed to produce a strong positive result to date, see, e.g., Refs.~\cite{SuperCDMS2014,CoGeNT2011,CRESST2014,DAMA2008,LUX2013,XENON2013,Roberts2016DAMA}, which has spurred significant interest of late in searching for alternate well-motivated forms of dark matter, such as ultralight (sub-eV mass) spin-0 particles that form either an oscillating classical field or topological defects, see, e.g., Refs.~\cite{ADMX2010,Graham2011,Baker2012,GNOME2013A,Rubakov2013,Beck2013,Sikivie2014LC,Stadnik2014axions,CASPER2014,Roberts2014prl,Porayko2014,Sikivie2014Atomic,Stadnik2014defects,Derevianko2014,Roberts2014long,Arvanitaki2015scalar,Stadnik2015laser,Beck2015,Popov2015,Budker2015scalar,Stadnik2015DM-VFCs,Bize2016scalar,Stadnik2016scalar}. 

The idea that the fundamental constants of Nature might vary with time can be traced as far back as the large numbers hypothesis of Dirac, who hypothesised that the gravitational constant $G$ might be proportional to the reciprocal of the age of the Universe \cite{Dirac1937}. More contemporary theories, which predict a variation of the fundamental constants on cosmological timescales, typically invoke a (nearly) massless underlying dark energy-type field, see, e.g., the review \cite{Uzan2011Review} and the references therein.
Most recently, a new model for the cosmological evolution of the fundamental constants of Nature has been proposed in Ref.~\cite{Stadnik2015DM-VFCs}, in which the interaction of an oscillating classical scalar dark matter field with ordinary matter via quadratic interactions produces both `slow' linear-in-time drifts and oscillating-in-time variations of the fundamental constants \cite{Footnote1coherence}.
Topological defects, which are stable, extended-in-space forms of dark matter that consist of light scalar dark matter fields stabilised by a self-interaction potential \cite{Vilenkin1985} and which interact with ordinary matter, produce transient-in-time 
variations of the fundamental constants \cite{Derevianko2014,Stadnik2014defects}. 
The oscillating-in-time and transient-in-time variations of the fundamental constants produced by scalar dark matter can be sought for in the laboratory using high-precision measurements, which include atomic clocks \cite{Prestage1995clock,Flambaum1999clock,Angstmann2004clock,Fischer2004clock,Flambaum2006clock,Rosenband2008clock,Budker2013clock,Gill2014clock,Peik2014clock,Katori2015clock}, highly-charged ions \cite{Labzowsky2005HCI,Berengut2010HCI,Berengut2011HCI,Berengut2012HCI,Derevianko2012HCI,Safronova2014HCI,Windberger2015HCI}, molecules \cite{Korobov2005mol,Flambaum2006mol,Hudson2006mol,Flambaum2007mol,Ye2008mol,DeMille2008mol} and nuclear clocks \cite{Flambaum2006nuclear,Friar2007nuclear,Ren2008nuclear,Flambaum2009nuclear,Berengut2009nuclear,Litvinova2009nuclear,Hudson2009nuclear}, in which two transition frequencies are compared over time. 

Instead of comparing two transition frequencies over time, we may instead compare a photon wavelength with an interferometer arm length, in order to search for variations of the  fundamental constants \cite{Stadnik2015laser,Stein1976,Tobar2010} (see also \cite{Schiller2004,Tobar2015} for some other applications). 
In the present work, we outline new laser interferometer measurements to search for variation of the electromagnetic fine-structure constant and particle masses (including a non-zero photon mass). 
We propose a strontium optical lattice clock -- silicon single-crystal cavity interferometer as a novel small-scale platform for these new measurements. 
The small-scale hydrogen maser -- cryogenic sapphire oscillator system \cite{Tobar2010} and large-scale gravitational-wave detectors, such as LIGO-Virgo \cite{LIGO-Virgo2016}, GEO600 \cite{GEO600_2011Update}, TAMA300 \cite{TAMA300_2008Update}, eLISA \cite{eLISA2012Proposal} or the Fermilab Holometer \cite{FL_Holometer_2009Proposal}, can also be used as platforms for some of our newly proposed measurements. 

\section{General Theory}
\label{Sec:Theory}
Unless explicitly indicated otherwise, we employ the natural units $\hbar = c = 1$ in the present work.
Alterations in the electromagnetic fine-structure constant $\alpha = e^2/\hbar c$, where $-e$ is the electron charge, $\hbar = h/2\pi$ is the reduced Planck constant and $c$ is the photon speed, or particle masses (including a non-zero photon mass $m_\gamma$) produce alterations in the accumulated phase of the light beam inside an interferometer $\Phi  = \omega L / c$, since an atomic transition frequency $\omega$ and length of a solid $L \sim N a_{\textrm{B}}$, where $N$ is the number of atoms and $a_{\textrm{B}} = \hbar^2/m_e e^2$ is the Bohr radius ($m_e$ is the electron mass), both depend on the fundamental constants and particle masses. Alterations in the accumulated phase can be expressed in terms of the sensitivity coefficients $K_X$, which are defined by:
\begin{equation}
\label{Sens_coeffns_defn}
\frac{\delta \Phi }{\Phi} = \sum_{X_i = \alpha, m_e, ...} K_{X_i} \frac{\delta X_i}{X_i} + K_{m_\gamma} \left( \frac{m_\gamma}{m_e} \right)^2  , 
\end{equation}
where the sum runs over all relevant fundamental constants $X_i = \alpha, m_e, ...$ (except photon mass). 
The sensitivity coefficients depend on the specific measurement that is performed. 
In order to define the variation of dimensionful parameters, such as $m_e$, we assume that such variations are due to the interactions of dark matter with ordinary matter, see, e.g., Ref.~\cite{Stadnik2015DM-VFCs}.
The sensitivity coefficients, which we derive below in Sections \ref{Sec:3} and \ref{Sec:4}, are for single-arm interferometers, but are readily carried over to the case of two-arm Michelson-type interferometers, for which the observable quantity is the phase difference $\Delta \Phi = \Phi_1 - \Phi_2$ between the two arms, as we illustrate with a couple of examples in Section \ref{Sec:5}.

One intuitively expects that multiple reflections should enhance observable effects due to variation of the fundamental constants by the effective mean number of passages $N_{\textrm{eff}}$. 
This can be readily verified by the following simple derivation. 
For multiple reflections of a continuous light source that forms a standing wave (in the absence of variation of the fundamental constants), we sum over all possible number of reflections $n$:
\begin{equation}
\label{App-A_derivation1}
\sum_{n=1}^{\infty}  \exp \left[ - n \left( \kappa - i  \Phi \right)  \right] = \frac{1}{\exp \left( \kappa - i \Phi \right)  - 1} ,
\end{equation}
where $\kappa \equiv 1/N_{\textrm{eff}}$ is the attenuation factor that accounts for the loss of light amplitude after a single to-and-back passage along the length of the arm, and $\Phi = 2 \pi m + \delta \Phi$ ($m$ is an integer) is the phase accumulated by the light beam in a single to-and-back passage along the length of the arm. For a large effective mean number of passages, $N_{\textrm{eff}} \gg 1$, and for sufficiently small deviations in the accumulated phase, $N_{\textrm{eff}} \cdot \delta \Phi \ll 1$, the sum in Eq.~(\ref{App-A_derivation1}) can be written as: 
\begin{align}
\label{App-A_derivation2}
\sum_{n=1}^{\infty}  \exp \left[ - n \left( \kappa - i \Phi \right)  \right] &\simeq  N_{\textrm{eff}} \exp \left(i N_{\textrm{eff}} \cdot \delta \Phi \right)  ,
\end{align}
from which it is evident 
that the effects of small variations in the accumulated phase are enhanced by the factor $N_{\textrm{eff}}$.

\section{Variation of the electromagnetic fine-structure constant and particle masses}
\label{Sec:3}
Variation of $\alpha$ and particle masses alters the accumulated phase through alteration of $\omega$ and $L \sim N a_{\textrm{B}}$.
There are four main classes of experimental configurations to consider, depending on whether the frequency of light inside an interferometer is determined by a specific atomic transition (i.e., when the high-finesse cavity length is stabilised to an atomic transition) or by the length of a resonator (i.e., when the laser is stabilised to a high-finesse cavity), and whether the interferometer arm length is allowed to vary freely (i.e., allowed to depend on the length of the solid spacer between the mirrors) or its fluctuations are deliberately shielded (i.e., the arm length is made independent of the length of the solid spacer between the mirrors, e.g., through the use of a multiple-pendulum mirror system). We consider each of these configurations in turn.

\subsection{Configuration A (atomic transition frequency, free arm length)} 
The simplest case is when the frequency of light inside an interferometer is determined by an optical atomic transition frequency and the interferometer arm length is allowed to vary freely (i.e., allowed to depend on the length of the solid spacer between the mirrors). 
A strontium clock -- silicon cavity interferometer in its standard mode of operation falls into this category. 
In this case, the atomic transition wavelength and arm length are compared directly:
\begin{align}
\label{Phase_accumulated_1}
\Phi = \frac{\omega L}{c} \propto \left(\frac{e^2}{a_{\textrm{B}} \hbar} \right) \left(\frac{N a_{\textrm{B}}}{c} \right) = N \alpha , 
\end{align}
where the optical atomic transition frequency $\omega$ is proportional to the atomic unit of frequency $e^2/a_{\textrm{B}} \hbar$.
Variation of $\alpha$ thus gives rise to the following phase shift:
\begin{align}
\label{Sensitivity_coefficient_1}
\delta\Phi \simeq \Phi \frac{\delta \alpha}{\alpha} .
\end{align}
We note that the effect of variation of $\alpha$ already appears at the non-relativistic level in Eq.~(\ref{Sensitivity_coefficient_1}), with the corresponding sensitivity coefficient 
$K_\alpha = 1$. 
For systems consisting of light elements, the relativistic corrections to this sensitivity coefficient are small and can be neglected. 
This is in stark contrast to optical clock comparison experiments, for which $K_\alpha = 0$ in the non-relativistic approximation and the contributions to $K_\alpha$ arise solely from relativistic corrections \cite{Flambaum1999clock,Angstmann2004clock}.

For a strontium clock -- silicon cavity interferometer, which operates on the $^{87}$Sr $^{1}S_0$ $-$ $^{3}P_0$ transition ($\lambda = 698$ nm) and for which the cavity length is $L = 0.21$ m \cite{Ye2012}, the phase shift in Eq.~(\ref{Sensitivity_coefficient_1}) for a single to-and-back passage of the light beam is:
\begin{align}
\label{Sensitivity_coefficient_1a}
\delta \Phi \simeq 3.8 \times 10^6 \frac{\delta \alpha}{\alpha} .
\end{align}
For comparison, in a large-scale gravitational-wave detector of length $L = 4$ km and operating on a typical atomic optical transition frequency, the phase shift for a single to-and-back passage of the light beam is:
\begin{align}
\label{Sensitivity_coefficient_1b}
\delta \Phi \sim 10^{11} \frac{\delta \alpha}{\alpha} .
\end{align}
As noted in Section \ref{Sec:Theory}, multiple reflections enhance the coefficients in Eqs.~(\ref{Sensitivity_coefficient_1a}) and (\ref{Sensitivity_coefficient_1b}) by the effective mean number of passages $N_{\textrm{eff}}$, which depends on the reflectivity properties of the mirrors used. 
For large-scale interferometers, this enhancement factor is $N_{\textrm{eff}} \sim 10^2$. 
For small-scale interferometers with highly-reflective mirrors, this enhancement factor can be considerably larger:~$N_{\textrm{eff}} \sim 10^5$.

Another possible system in this category is the hydrogen maser -- cryogenic sapphire oscillator system, which operates on the $^{1}$H ground state hyperfine transition: 
\begin{equation}
\label{omega_H_hf}
\omega \propto \left(\frac{e^2}{a_{\textrm{B}} \hbar}\right) \left[\alpha^2 F_{\textrm{rel}}(Z\alpha)\right] \left[ \mu_p \frac{m_e}{m_p} \right] ,
\end{equation}
where $F_{\textrm{rel}}(Z\alpha) \simeq 1$ is the relativistic Casimir factor and $\mu_p$ is the dimensionless magnetic dipole moment of the proton in units of the nuclear magneton.
In this case, changes in the measured phase have the following dependence on changes in the fundamental constants:
\begin{equation}
\label{Sensitivity_coefficient_1CSO}
\frac{\delta \Phi}{\Phi} \simeq  3 \frac{\delta \alpha}{\alpha} + \frac{\delta m_e}{m_e} - 0.14 \frac{\delta m_q}{m_q} , 
\end{equation}
where $m_q = (m_u + m_d) / 2$ is the averaged light quark mass, and where we have used the calculated values $\delta \mu_p / \mu_p = -0.09  \delta m_q / m_q$ \cite{Flambaum2004Thomas} and $\delta m_p / m_p = +0.05  \delta m_q / m_q$ \cite{Flambaum2004Thomas,Flambaum2006Roberts}.

If one performs two simultaneous interferometry experiments with two different transition lines, using the same set of mirrors, then one may search for variations of the fundamental constants associated with changes in the atomic transition frequencies:
\begin{equation}
\label{seismic_shield}
\delta X = \frac{c (\omega_A \delta \Phi_B - \omega_B \delta \Phi_A)}{L (\omega_A \frac{\partial \omega_B}{\partial X} - \omega_B \frac{\partial \omega_A}{\partial X})} .
\end{equation}
In particular, note that shifts in the arm lengths (due to variation of the fundamental constants or undesired effects, such as seismic noise or tidal effects) cancel in Eq.~(\ref{seismic_shield}). 
We also note that atomic clock transition frequencies may also be compared by locking lasers to the atomic transitions and using phase coherent optical mixing and frequency comb techniques to measure the laser frequency difference/ratio.

\subsection{Configuration B (atomic transition frequency, fixed arm length)} 
If fluctuations in the arm length are deliberately shielded (i.e., the arm length is made independent of the length of the solid spacer between the mirrors, e.g., through the use of a multiple-pendulum mirror system), but $\omega$ is still determined by an atomic transition frequency, then changes in the measured phase $\Phi \propto \omega/c \propto m_e e^4 / \hbar^3 c = (m_e c / \hbar) \cdot (e^2 / \hbar c)^2$ have the following dependence on changes in the fundamental constants:
\begin{equation}
\label{Sensitivity_coefficient_4}
\frac{\delta \Phi}{\Phi} \simeq \frac{\delta m_e}{m_e} + 2 \frac{\delta \alpha}{\alpha} .
\end{equation}

\subsection{Configuration C (resonator-determined wavelength, free arm length)} 
When a laser is locked to a resonator mode determined by the length of the resonator, $\omega$ is determined by the length of the resonator, which changes if the fundamental constants change. In the non-relativistic limit, the wavelength and arm length (as well as the size of Earth) have the same dependence on the Bohr radius, and so there are no observable effects if changes of the fundamental constants are slow (adiabatic) and if the interferometer arm length is allowed to vary freely (i.e., allowed to depend on the length of the solid spacer between the mirrors). 
Indeed, this may be viewed as a simple change in the measurement units. Transient effects due to the passage of topological defects may still produce effects, since changes in $\omega$ and $L$ may occur at different times.

The sensitivity of laser interferometry to non-transient effects is determined by relativistic corrections, which we estimate as follows. The size of an atom $R$ is determined by the classical turning point of an external atomic electron. Assuming that the centrifugal term $\sim 1/R^2$ is small at large distances, we obtain  $(Z_i+1)e^2/R = -E$, where $E$ is the energy of the external electron and $Z_i $ is the net charge of the atomic species (for a neutral atom, $Z_i=0$). This gives the relation: $\delta R/ R = \delta (E/e^2) / |E/e^2|$. 
The single-particle relativistic correction to the energy in a many-electron atomic species is given by \cite{Flambaum1999clock}:
\begin{equation}
\label{rel-correxn_FS}
\Delta_n \simeq E_n \frac{(Z\alpha)^2}{\nu (j+1/2)} ,
\end{equation}
where $E_n = - m_e e^4 (Z_i+1)^2 / 2 \hbar^2 \nu^2$ is the energy of the external atomic electron, 
$j$ is its angular momentum, $Z$ is the nuclear charge, and $\nu \sim 1$ is the effective principal quantum number. Variation of $\alpha$ thus gives rise to the following phase shift:
\begin{equation}
\label{Sensitivity_coefficient_2}
\frac{\delta \Phi}{\Phi} \simeq 2\alpha^2 \left[\frac{Z_{\textrm{res}}^2}{\nu_{\textrm{res}} (j_{\textrm{res}} + 1/2)} - \frac{Z_{\textrm{arm}}^2}{\nu_{\textrm{arm}} (j_{\textrm{arm}} + 1/2)} \right]  \frac{\delta \alpha}{\alpha} . 
\end{equation}
Here $Z_{\textrm{res}}$ is the atomic number of the atoms that make up the solid spacer between the mirrors of the resonator, while $Z_{\textrm{arm}}$ is the atomic number of the atoms that make up the arm. 
Note that the sensitivity coefficient depends particularly strongly on the factor $Z^2$. $\left|K_\alpha\right| \ll 1$ for light atoms and may be of the order of unity in heavy atoms.

\subsection{Configuration D (resonator-determined wavelength, fixed arm length)} 
If fluctuations in the arm length are deliberately shielded (i.e., the arm length is made independent of the length of the solid spacer between the mirrors) and $\omega$ is determined by the length of the resonator, then changes in the measured phase $\Phi \propto 1/\lambda \propto 1/a_{\textrm{B}}$ have the following dependence on changes in the fundamental constants:
\begin{equation}
\label{Sensitivity_coefficient_3}
\frac{\delta \Phi}{\Phi} \simeq \frac{\delta m_e}{m_e} + \frac{\delta \alpha}{\alpha} .
\end{equation}
A large-scale gravitational-wave detector (such as LIGO, Virgo, GEO600 or TAMA300) in its standard mode of operation falls into this category.

\section{Non-zero photon mass} 
\label{Sec:4}
A non-zero photon mass alters the accumulated phase through alteration of $\omega$, $L = N R$ (where $R$ is the atomic radius) and $c$. In particular, if a non-zero photon mass is generated due to the interaction of photons with slowly moving dark matter ($v_{\textrm{DM}} \ll 1$), then the energy and momentum of the photons are approximately conserved and the photon speed changes according to:
\begin{equation}
\label{delta_c/c}
\delta c \simeq - \frac{m_\gamma^2 }{2  \omega^2 } .
\end{equation}

The effects of a non-zero photon mass in atoms are more subtle. The potential of an atomic electron changes from Coulomb to Yukawa-type:
\begin{align}
\label{Coulomb_to_Yukawa}
&V_{\textrm{Coulomb}} (r) = \sum_{i} \frac{e^2}{|\v{r} - \v{r}_i|} - \frac{Ze^2}{r} , \\
=> ~  &V_{\textrm{Yukawa}} (r) = \sum_{i} \frac{ e^{-m_\gamma |\v{r} - \v{r}_i| } e^2 }{|\v{r} - \v{r}_i|} - \frac{ e^{-m_\gamma r} Ze^2 }{r} ,
\end{align}
where the sum runs over all remaining atomic electrons.
For $m_\gamma r  \ll 1$, the leading term of the corresponding perturbation reads (we omit the constant terms, which do not alter the atomic transition frequencies and wavefunctions):
\begin{equation}
\label{Yukawa_perturbation}
\delta V (r) =  \frac{e^2 m_\gamma^2 }{2 } \left[ \sum_{i}{|\v{r} - \v{r}_i|} - Zr  \right] ,  
\end{equation}
which for a neutral atom takes the asymptotic forms:
\begin{eqnarray}
\label{Yukawa_perturbation_asymptotic}
\delta V (r) \simeq \left\{ \begin{array}{ll}
- Z e^2 m_\gamma^2  r /2 & \textrm{when $r \ll a_{\textrm{B}}/Z^{1/3}$,}\\
- e^2 m_\gamma^2 r /2  & \textrm{when $r \gg a_{\textrm{B}}/Z^{1/3}$.}
\end{array} \right.
\end{eqnarray}
In the semiclassical approximation, it is straightforward to confirm that the dominant contribution to the expectation value of the operator (\ref{Yukawa_perturbation}) comes from large distances, $r \gg a_{\textrm{B}}/Z^{1/3}$, where the external electron sees an effective charge of $Z_{\textrm{eff}} = 1$. Therefore, the shift in an atomic energy level $A$ is simply:
\begin{equation}
\label{Yukawa_energy_shift}
\delta E_A \simeq - \frac{e^2 m_\gamma^2  R_A}{2} ,
\end{equation}
where $R_A = \left< A \left| r \right|A \right> $ is the expectation value of the radius operator for state $A$. Typically, $R_A \sim \textrm{several} ~ a_{\textrm{B}}$.
Assuming that the perturbation (\ref{Yukawa_perturbation}) is adiabatic and that the the dominant contribution to the matrix elements $\left< n \left|  \delta V \right|A \right>$ comes from large distances, application of time-independent perturbation theory gives the following shift in the size of the atomic orbit for state $A$:
\begin{align}
\label{delta_R_Yukawa}
\delta R_A \simeq -m_\gamma^2 \sum_{n \ne A} \frac{\left< A \left| er \right|n \right> \left< n \left| er \right|A \right>}{E_{A}^{(0)} - E_{n}^{(0)}} \sim m_\gamma^2 \alpha_A ,
\end{align}
where $\alpha_A$ is the static dipole polarisability of state $A$. Static dipole polarisabilities for the electronic ground states of neutral atoms range from $4.5$ $a_{\textrm{B}}^3$ in hydrogen to $400$ $a_{\textrm{B}}^3$ in caesium \cite{Schwerdtfeger2014Table}.


\subsection{Configuration A (atomic transition frequency, free arm length)} 
If $\omega$ is determined by an atomic transition frequency and the interferometer arm length is allowed to vary freely (i.e., allowed to depend on the length of the solid spacer between the mirrors), then a non-zero photon mass produces the following changes in the measured phase $\Phi = \omega L/c$:
\begin{align}
\label{Sensitivity_coefficient_1B}
\frac{\delta \Phi}{\Phi} &\simeq  \frac{e^2 m_\gamma^2 (R_{f} - R_{i} ) }{2 \omega} + \frac{m_\gamma^2 \alpha_{\textrm{arm}} }{R_{\textrm{arm}} } + \frac{m_\gamma^2}{2 \omega^2}  \simeq \frac{m_\gamma^2}{2 \omega^2} ,
\end{align}
where $R_{f} - R_{i} = \left< f \left| r \right|f \right> - \left< i \left| r \right|i \right>$ is the difference in the orbital size between the final and initial states involved in the radiative atomic transition, and $\alpha_{\textrm{arm}}$ is the static dipole polarisability of the atoms that make up the arm. 
The three separate contributions in Eq.~(\ref{Sensitivity_coefficient_1B}) scale roughly in the ratio ${\alpha^2:\alpha^2:1}$, respectively, meaning that the contribution from the change in the photon speed dominates.

\subsection{Configuration B (atomic transition frequency, fixed arm length)} 
If fluctuations in the arm length are deliberately shielded (i.e., the arm length is made independent of the length of the solid spacer between the mirrors), but $\omega$ is still determined by an atomic transition frequency, then a non-zero photon mass produces the following changes in the measured phase $\Phi \propto \omega/c$:
\begin{align}
\label{Sensitivity_coefficient_4B}
\frac{\delta \Phi}{\Phi} &\simeq  \frac{e^2 m_\gamma^2  (R_{f} - R_{i} ) }{2 \omega} + \frac{m_\gamma^2}{2 \omega^2}  \simeq \frac{m_\gamma^2}{2 \omega^2} ,
\end{align}
where we again note that the contribution from the change in the photon speed dominates.

\subsection{Configuration C (resonator-determined wavelength, free arm length)} 
If $\omega$ is determined by the length of the resonator and the interferometer arm length is allowed to vary freely (i.e., allowed to depend on the length of the solid spacer between the mirrors), then a non-zero photon mass produces the following changes in the measured phase $\Phi = 2 \pi L / \lambda$:
\begin{equation}
\label{Sensitivity_coefficient_2B}
\frac{\delta \Phi}{\Phi} \sim  m_\gamma^2 \left( \frac{\alpha_{\textrm{arm}} }{R_{\textrm{arm}}} - \frac{\alpha_{\textrm{res}} }{R_{\textrm{res}}} \right)  .
\end{equation}
Here $\alpha_{\textrm{res}}$ is the static dipole polarisability of the atoms that make up the solid spacer between the mirrors of the resonator. 
The phase shift in Eq.~(\ref{Sensitivity_coefficient_2B}) is suppressed by the factor $\sim \alpha^2$ in the static limit (compare with Eqs.~(\ref{Sensitivity_coefficient_1B}) and (\ref{Sensitivity_coefficient_4B}) above). However, for time-dependent effects, the phase shift can be significantly larger (see the examples in Section \ref{Sec:5}).

\subsection{Configuration D (resonator-determined wavelength, fixed arm length)} 
If fluctuations in the arm length are deliberately shielded (i.e., the arm length is made independent of the length of the solid spacer between the mirrors) and $\omega$ is determined by the length of the resonator, then a non-zero photon mass produces the following changes in the measured phase $\Phi \propto 1/\lambda$:
\begin{equation}
\label{Sensitivity_coefficient_3B}
\frac{\delta \Phi}{\Phi} \sim  - m_\gamma^2  \frac{\alpha_{\textrm{res}} }{R_{\textrm{res}}}  .
\end{equation}
Similarly to Eq.~(\ref{Sensitivity_coefficient_2B}), the phase shift in Eq.~(\ref{Sensitivity_coefficient_3B}) is also suppressed by the factor $\sim \alpha^2$ in the static limit. However, we again note that the phase shift can be significantly larger for time-dependent effects (see Section \ref{Sec:5}).

\section{Specific examples}
\label{Sec:5}
\subsection{Oscillating classical dark matter (effects of spatial coherence)}
Oscillating classical dark matter exhibits not only temporal coherence \cite{Footnote1coherence}, but also spatial coherence, with a coherence length given by: $l_{\textrm{coh}} \sim 2\pi / m_\phi v_{\textrm{vir}} \sim 10^3 \cdot 2\pi / m_\phi$, where $m_\phi$ is the dark matter particle mass, and a virial (root-mean-square) speed of $v_{\textrm{vir}} \sim 10^{-3} $ is typical in our local galactic neighbourhood. Our Solar System travels through the Milky Way (and hence relative to galactic dark matter) at a comparable speed $\left< v \right> \sim v_{\textrm{vir}} \sim 10^{-3} $. An oscillating scalar dark matter field takes the form:
\begin{equation}
\label{Coh_spat_DM_field}
\phi \left(\v{r}, t\right) \simeq \phi_0 \cos \left( m_\phi t -  m_\phi \left<\v{v}\right> \cdot \v{r} \right) ,
\end{equation}
meaning that measurements performed on length scales $l \lesssim 2\pi / m_\phi v_{\textrm{vir}}$ are sensitive to dark matter-induced effects that arise from differences in the spatial phase term $m_\phi \left<\v{v}\right> \cdot \v{r}$ at two or more points.

\begin{figure}[h!]
\begin{center}
\includegraphics[width=6cm]{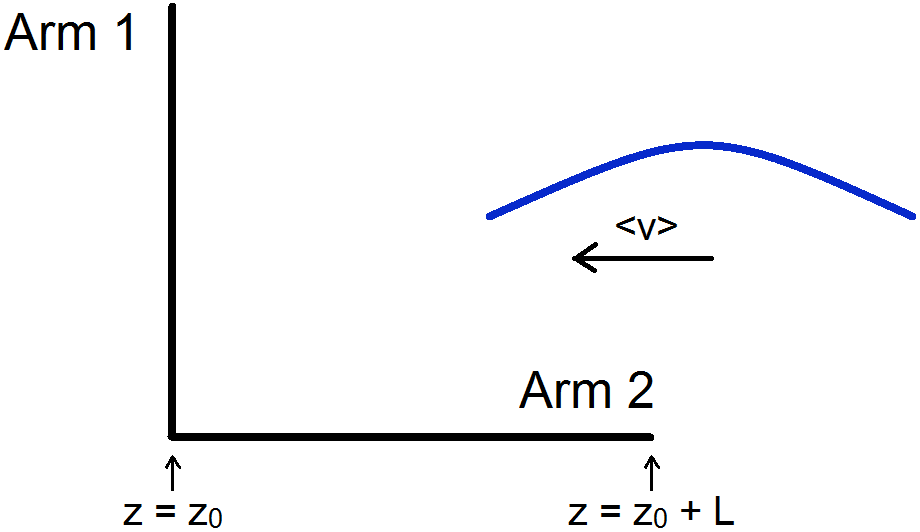}
\caption{(Color online) Passage of dark matter directly onto an arm of a gravitational-wave detector ($L_1 = L_2 = L$). 
} 
\label{fig:Interferometer_spatial_coherence}
\end{center}
\end{figure}

As a specific example, we consider measurements performed using a large-scale gravitational-wave detector with equal arm lengths that are deliberately shielded from fluctuations, $L_1 = L_2 = L = \textrm{constant}$, and with the emitted photon wavelength determined by the length of the resonator. 
Since we are considering slowly moving dark matter ($v_{\textrm{DM}} \ll 1$), changes in the wavelength of the travelling photon are related to changes in $c$ by: $\delta \lambda / \lambda \simeq \delta c  \simeq - [m_\gamma(\v{r},t)]^2 / 2 \omega^2$, where the interaction between the photon field and $\phi^2$ may be interpreted as the varying photon mass: $[m_\gamma \left(\v{r}, t\right)]^2 = (m_\gamma)_{\textrm{max}}^2 {\cos^2 \left( m_\phi t -  m_\phi \left<\v{v}\right> \cdot \v{r} \right)} $.  
For the simplest case when the dark matter is incident directly onto one of the detector arms as shown in Fig.~\ref{fig:Interferometer_spatial_coherence}, the shift in the accumulated phase difference between the two arms is given by:
\begin{equation}
\label{LIGO_example_A}
\delta \left( \Phi_1 - \Phi_2 \right) = \frac{2\pi}{\lambda} \int_{z_0}^{z_0 + L} \left[ \frac{\delta \lambda (z)}{\lambda} - \frac{\delta \lambda (z_0)}{\lambda} \right] dz ,
\end{equation}
and to leading order we find:
\begin{equation}
\label{LIGO_example_B}
\frac{\delta  \left( \Phi_1 - \Phi_2 \right)}{\Phi} \simeq \frac{(m_\gamma)_{\textrm{max}}^2 m_\phi \left< v \right> L}{4 \omega^2} \sin \left( 2 m_\phi t + 2 m_\phi \left< v \right> z_0  \right) .
\end{equation}
The shift in the accumulated phase difference between the two arms in Eq.~(\ref{LIGO_example_B}) is suppressed by the factor $m_\phi \left< v \right> L < 1$.


\subsection{Topological defect dark matter}
Topological defect dark matter is intrinsically coherent, both temporally and spatially. As a specific example, we again consider measurements performed using a large-scale gravitational-wave detector with equal arm lengths that are deliberately shielded from fluctuations and with the emitted photon wavelength determined by the length of the resonator. For the case of a 2D domain wall with a Gaussian cross-sectional profile of root-mean-square width $d \sim 1/m_\phi$ and which travels slowly ($v_{\textrm{TD}} \ll 1$) in the geometry shown in Fig.~\ref{fig:Interferometer_spatial_coherence}, the interaction between the photon field and $\phi^2$ may be interpreted as the varying photon mass: $[m_\gamma \left(z, t\right)]^2 = (m_\gamma)_{\textrm{max}}^2 {\exp[-(z + vt)^2 / d^2]} $.  
Calculating the shift in the accumulated phase difference between the two arms, Eq.~(\ref{LIGO_example_A}), we find to leading order:
\begin{align}
\label{LIGO_example_C}
&\frac{\delta  \left( \Phi_1 - \Phi_2 \right)}{\Phi} \simeq \frac{(m_\gamma)_{\textrm{max}}^2 }{2 \omega^2} \left\{ \exp\left[- \frac{\left(z_0 + tv \right)^2}{d^2}\right]  \right. \notag \\
 &- \left.  \frac{\sqrt{\pi}d}{2L} \left[\textrm{erf} \left( \frac{L + tv +z_0}{ d} \right) - \textrm{erf} \left( \frac{tv +z_0}{ d} \right)  \right] \right\}
,
\end{align}
where erf is the standard error function, defined as $\textrm{erf}(x) = \left( 2/\sqrt{\pi} \right) \int_0^x e^{-u^2} du$.
The shift in the accumulated phase difference between the two arms in Eq.~(\ref{LIGO_example_C}) is largest for $d \sim L$. For $d \gg L$, the phase shift in (\ref{LIGO_example_C}) is suppressed by the factor $L/d \ll 1$. In the case when $d \ll L$, the phase shift in (\ref{LIGO_example_C}) is suppressed by the factor $d/L \ll 1$ when the topological defect envelops arm 2 but remains far away from arm 1; however, at the times when the topological defect envelops arm 1, there is no such suppression.

\section{Conclusions}
We have outlined new laser interferometer measurements to search for variation of the electromagnetic fine-structure constant $\alpha$ and particle masses (including a non-zero photon mass). 
We have proposed a strontium optical lattice clock -- silicon single-crystal cavity interferometer as a novel small-scale platform for these new measurements. 
Our proposed laser interferometer measurements, which may also be performed with large-scale gravitational-wave detectors, such as LIGO, Virgo, GEO600 or TAMA300, may be implemented as an extremely precise tool in the direct detection of scalar dark matter.
For oscillating classical scalar dark matter, a single interferometer is sufficient in principle, while for topological defects, a global network of interferometers is required.
The possible range of frequencies for oscillating classical dark matter is $10^{-8}~\textrm{Hz} \lesssim f \lesssim 10^{13}~\textrm{Hz}$ (corresponding to the dark matter particle mass range $10^{-22}~\textrm{eV} \lesssim m_\phi \lesssim 0.1~\textrm{eV}$), while the timescale of passage of topological defects through a global network of detectors is $T \sim R_{\textrm{Earth}} / v_{\textrm{TD}} \sim 20$ s for a typical defect speed of $v_{\textrm{TD}} \sim 300$ km/s. 
The current best sensitivities to length fluctuations are at the fractional level $\sim 10^{-22} - 10^{-23}$ in the frequency range $\sim 20 - 2000$ Hz for a large-scale gravitational-wave detector \cite{LIGO-Virgo2016} and at the fractional level $\sim 10^{-15} - 10^{-16}$ in the frequency range $\sim 0.01 - 10$ Hz for a silicon-based cavity \cite{Ye2012}.


\section*{ACKNOWLEDGEMENTS}
We are very grateful to Jun Ye and Fritz Riehle for suggesting the strontium clock -- silicon cavity interferometer as a suitable small-scale platform for our newly proposed measurements, and for important discussions. 
We are also grateful to an anonymous referee for suggesting the use of phase coherent optical mixing and frequency comb techniques, further to our proposal centred around Eq.~(\ref{seismic_shield}). 
We would like to thank Bruce Allen, Dmitry Budker, Federico Ferrini, Hartmut Grote, Sergey Klimenko, Giovanni Losurdo, Guenakh Mitselmakher and Surjeet Rajendran for helpful discussions. 
This work was supported by the Australian Research Council. V.~V.~F.~is grateful to the Mainz Institute for Theoretical Physics (MITP) for its hospitality and support.




\end{document}